\documentstyle[epsfig,preprint,aps]{revtex}
\begin{document}
\tighten
\preprint{
\vbox{
\hbox{ADP-01-06/T441}
}}
\draft
\title{Pions in isospin asymmetric nuclei}
%
\author{K. Saito$^a$\footnote{
E-mail addresses: 
ksaito@nucl.phys.tohoku.ac.jp}, 
V. Guzey$^b$\footnote{vguzey@physics.adelaide.edu.au}, 
K. Tsushima$^b$\footnote{ktsushim@physics.adelaide.edu.au} 
and A.W. Thomas$^b$\footnote{athomas@physics.adelaide.edu.au}}
\address{$^a$ Tohoku College of Pharmacy, Sendai 981-8558, Japan \\
$^b$ Department of Physics and Mathematical Physics and \\
Special Research Center for the Subatomic Structure of Matter (CSSM) \\
University of Adelaide, Adelaide SA 5005, Australia}
\maketitle
\begin{abstract}
Using a pair of the lightest mirror nuclei, $^3$He and $^3$H, we study 
the effect of the medium modification of pion fields on the flavor 
non-singlet structure function.  The change of the pion fields leads to 
an enhancement of the flavor asymmetry of the 
antiquark distributions in a nucleus. 
\end{abstract}

\vspace{1cm}

The partonic distribution functions of the nucleon, in particular the 
flavor dependence of the antiquark distributions, are of considerable 
interest~\cite{review1}. 
Within the framework of perturbative QCD (pQCD) the light quark sea is 
expected to be flavor symmetric. However, the experimental 
data~\cite{exp1} contradict this idea, revealing 
an excess of ${\bar d}$ over ${\bar u}$ in the free proton. 
This inconsistency indicates that non-perturbative effects 
should be responsible for the flavor asymmetry in the light sea quark  
distributions. 
For example, some flavor asymmetry was anticipated before 
the measurements on the basis 
of the chiral structure of the nucleon~\cite{pion1}. 

The physical proton has a relatively large $\pi^+$-neutron 
Fock component which naturally leads to a
surplus of ${\bar d}$~\cite{review1,Field}.  
It is known that this Fock component 
offers the main contribution to the ${\bar d}$ excess
and that the contributions of the other mesons and $\Delta$ isobars
have opposite signs and tend to cancel each other~\cite{review1}.
An alternative explanation for an excess of $\bar{d}$ over $\bar{u}$
involved the Pauli exclusion principle, given that there are two valence
$u$ quarks in the proton and one valence $d$~\cite{Field}.
Perturbative estimates failed to support this~\cite{Ross}, 
with the first non-perturbative explanation of the origin of such an effect 
in terms of the vacuum structure of the proton given by 
Signal and Thomas~\cite{SignalP}. Estimates of this effect within 
chiral quark models have also been given in 
Refs.~\cite{Wakamatsu,Diakonov,Goeke}.
It may well be that the experimentally observed excess involves 
contributions from both of these effects~\cite{MST}.

One way to learn more about the non-perturbative structure of the nucleon is 
to study the non-singlet difference between 
the proton (p) and neutron (n) structure 
functions, for nucleons bound 
in a pair of mirror nuclei~\cite{saito,guzey}.  
In this case any discrepancy between 
theoretical predictions and observed data will indicate a modification of 
the non-perturbative mechanism giving rise to the flavor asymmetry in the 
free proton, in the nuclear medium. 
In particular, such a discrepancy would be 
a sensitive probe to study pions in nuclei.  In this paper we examine the 
effect of changes in the pion cloud on the non-singlet combination of nuclear 
structure functions, using the lightest pair of mirror nuclei, $^3$He and 
$^3$H.  

How is the pion field modified in a nucleus ?  To study it we  
concentrate here on only Fock states consisting of a ``bare'' nucleon and 
pion, and ignore nuclear binding, Fermi motion and 
shadowing/antishadowing effects for the moment.  Under these assumptions,
the structure functions of the proton and neutron in the 
nucleus $A$ are given by~\cite{review1} 
\begin{eqnarray}
F_2^{p/A} &=& z_{p/A} {\tilde F}_2^p + f_{\pi^0p/p/A} \otimes 
{\tilde F}_2^p + f_{\pi^0p/p/A} \otimes F_2^{\pi^0} 
+  f_{\pi^+n/p/A} \otimes {\tilde F}_2^n +  f_{\pi^+n/p/A} \otimes 
F_2^{\pi^+} ,  \label{pinA} \\
F_2^{n/A} &=& z_{n/A} {\tilde F}_2^n + f_{\pi^0n/n/A} \otimes
{\tilde F}_2^n + f_{\pi^0n/n/A} \otimes F_2^{\pi^0}
+  f_{\pi^-p/n/A} \otimes {\tilde F}_2^p +  f_{\pi^-p/n/A} \otimes
F_2^{\pi^-} ,  \label{ninA} 
\end{eqnarray}
where ${\tilde F}_2^{p(n)}$ is the structure 
function of a 'bare' proton (neutron).  The probability to find the 
'bare' proton (neutron) in the physical proton (neutron) in $A$ is 
denoted by the normalization constant, $z_{p(n)/A}$.  
The shorthand notation, $f_{MB/N/A} \otimes 
F_2^K$, stands for the convolution of the (light-cone) momentum 
distribution of the pion $M$ ($= \pi^-, \pi^0, \pi^+$) per 
$N$ ($= p, n$), $f_{MB/N/A}(y)$ ($B = p, n$), and the structure 
function of $K$, $F_2^K(x)$ ($K = B, M$)~\cite{review1}: 
\begin{eqnarray}
f_{MB/N/A} \otimes F_2^B(x)&=& \int_0^{1-x} dy f_{MB/N/A}(y) 
F_2^B \left( \frac{x}{1-y} \right),  \label{convB} \\
f_{MB/N/A} \otimes F_2^M(x)&=& \int_x^{1} dy f_{MB/N/A}(y)
F_2^M \left( \frac{x}{y} \right).  \label{convM} 
\end{eqnarray}
The nuclear structure function is then simply given by 
$F_2^A(x) = Z F_2^{p/A}(x) + N F_2^{n/A}(x)$, where $Z$ and $N$ are the 
numbers of protons and neutrons, respectively.  

We consider a pair of mirror nuclei: $A = Z + N \ (Z > N)$ (proton rich)
and $A' = Z' + N' \ (N' > Z')$ (neutron rich).
In a nucleus we can expect a significant difference in the positive and
negative pion light-cone momentum distributions.
Recently, Korpa and Dieperink have calculated the pion fields in asymmetric
nuclear matter~\cite{diep}, with a result which is consistent with 
the Drell-Yan 
experiment of Alde et al.~\cite{dy}. Their result suggests that 
the difference in the distributions 
basically comes from two factors.  One is the Pauli blocking
of the nucleon in the final state; in the proton rich nucleus $A$ 
which we consider here, the
emission of $\pi^-$ (from a neutron, creating a proton in the final state)
is more suppressed than $\pi^+$ emission.  The other effect is the
dressing of the pion propagator in matter~\cite{Ericson}, 
where the particle-hole self-energy dominates.  
Korpa and Dieperink find that the delta-hole 
contribution is minor and that the neutral pion field is not much altered 
in the nuclear medium~\cite{diep}.  In summary, their analysis suggests 
that in the proton rich nucleus $A$ the $\pi^+ (\pi^-)$ 
field is enhanced (reduced) as compared with that in the free nucleon, while
the $\pi^0$ field is not changed a great deal.

In a nucleus the Coulomb interaction may affect the shape of the pion 
momentum distribution and, of course, 
one cannot use isospin symmetry: 
\begin{equation}
f_{\pi^+n/p/A} \neq f_{\pi^-p/n/A'} \ \ \mbox{and} \ \ 
f_{\pi^-p/n/A} \neq f_{\pi^+n/p/A'}. 
\label{csb}
\end{equation}
However, for the reasons discussed above, we suppose 
that the $\pi^0$ distribution 
is not changed much in matter: 
\begin{equation}
f_{\pi^0p/p/A} = f_{\pi^0n/n/A} = 
f_{\pi^0p/p/A'} = f_{\pi^0n/n/A'} \equiv f_{\pi^0N} .  
\label{pi0}
\end{equation}
Here $f_{\pi^0N}$ is the $\pi^0$ distribution in the free nucleon, 
which is given by the Sullivan process~\cite{review1}
\begin{equation}
f_{\pi^0N}(y)
= \frac{g^2}{16\pi^2y(1-y)^2} \int_0^\infty dk_t^2
\frac{F_{\pi N}^2(s)}{(M_N^2 - s)^2} (k_t^2 + y^2 M_N^2),
\label{fpi0}
\end{equation}
with $g(=13)$ the $\pi-N$ coupling constant, $k_t^2$ the transverse
momentum squared of the pion and
\begin{equation}
s = \frac{m_\pi^2 + k_t^2}{y} + \frac{M_N^2 + k_t^2}{1-y} . 
\label{s}
\end{equation}
The free nucleon mass is denoted $M_N$ (0.94 GeV) and $m_\pi$ (0.138 GeV)
is the pion mass.  The form factor, $F_{\pi N}(s)$, is given
by~\cite{review1}
\begin{equation}
F_{\pi N}(s) = \exp \left[\frac{M_N^2 -s}{2\Lambda^2} \right],
\label{formf}
\end{equation}
with $\Lambda$ the cut off parameter.

We divide the pion distribution into two pieces: 
\begin{eqnarray}
f_{\pi^+n/p/A(A')}(y) &=& f_{\pi^+n/p}(y) + \delta f_{\pi^+/A(A')}(y) , 
\label{fpipa} \\
f_{\pi^-p/n/A(A')}(y) &=& f_{\pi^-p/n}(y) + \delta f_{\pi^-/A(A')}(y),  
\label{fpima} 
\end{eqnarray}
where $f_{\pi^+n/p(\pi^-p/n)}$ is the momentum distribution of 
$\pi^+(\pi^-)$ in the free proton (neutron).  
The nuclear many-body effects on the pion field in $A (A')$ 
are expressed by $\delta f_{M/A(A')}$.  

The normalization constants in Eqs.(\ref{pinA}) and (\ref{ninA}) 
can be related to those for the free nucleon.  For example, 
using Eqs.(\ref{pi0}) and (\ref{fpipa}) we find
\begin{equation}
z_{p/A} = 
 1 - \langle f_{\pi^0N} \rangle - \langle f_{\pi^+n/p} \rangle
- \langle \delta f_{\pi^+/A} \rangle \equiv z_N - \langle \delta f_{\pi^+/A}
\rangle , 
\label{zzz}
\end{equation}
where $z_N$ ($= 1 - \langle f_{\pi^0N} \rangle - \langle f_{\pi^+n/p} 
\rangle$) is the normalization constant for the free nucleon. 
Finally then the nucleon structure functions in those nuclei become:
\begin{eqnarray}
F_2^{p/A} &=& F_2^p - \langle \delta f_{\pi^+/A} \rangle {\tilde F}_2^p 
+ \delta f_{\pi^+/A} \otimes [{\tilde F}_2^n +  F_2^{\pi^+}] ,  
\label{pinA2} \\
F_2^{n/A} &=& F_2^n - \langle \delta f_{\pi^-/A} \rangle {\tilde F}_2^n 
+ \delta f_{\pi^-/A} \otimes [{\tilde F}_2^p + F_2^{\pi^-}] ,  
\label{ninA2} \\
F_2^{p/A'} &=& F_2^p - \langle \delta f_{\pi^+/A'} \rangle {\tilde F}_2^p
+ \delta f_{\pi^+/A'} \otimes [{\tilde F}_2^n +  F_2^{\pi^+}] ,
\label{pinAp2} \\
F_2^{n/A'} &=& F_2^n - \langle \delta f_{\pi^-/A'} \rangle {\tilde F}_2^n
+ \delta f_{\pi^-/A'} \otimes [{\tilde F}_2^p + F_2^{\pi^-}] ,
\label{ninAp2}
\end{eqnarray}
where $F_2^{p(n)}$ is the free proton (neutron) structure function.  

Now we study the lightest mirror nuclei: $A = ^3$He and 
$A' = ^3$H.  In $^3$He, the $\pi^+$ meson is generated 
from the proton and the final state is given by 
$^3$He $= 2p + n \to p + 2n + \pi^+$, where 
$\pi^+$ feels a repulsive force from the Coulomb interaction  
with the single $p$.
(We neglect two pion emissions in the final states, such as 
$2p + n \to 3n + 2\pi^+$, for which the probability is 
expected to be very small.) 
On the other hand, the $\pi^-$ meson is produced 
by the neutron, and the final state is $3p + \pi^-$, where the $\pi^-$ feels 
a strong attractive force due to the $3p - \pi^-$ interaction.  Thus, 
we expect 
that the Coulomb force between $3p$ and $\pi^-$ is about three 
times stronger than that between $p$ and $\pi^+$ in the former case.  
In $^3$H, the $\pi^-$ feels an attractive force in the final 
state $2p + n + \pi^-$, and the Coulomb force is twice as 
large as that for the $\pi^+$ in $^3$He.  On the other hand, the $\pi^+$ 
in $^3$H does not feel any Coulomb force because the final state consists of 
$3n + \pi^+$.  

In order to evaluate   Eqs.(\ref{pinA2})-(\ref{ninAp2}), 
we need to estimate the distributions, $\delta f_{M/A}(y)$, 
in $^3$He and $^3$H, individually. 
The Coulomb force may change the shape of the pion momentum
distribution. As discussed above, since it acts on $\pi^-$ as an attractive 
force in the nucleus, the wave function of the pion in coordinate space  
shrinks. This means that the pion gets a (relatively) higher momentum and 
the shape of the distribution in momentum space should 
shift toward larger $y$.  For the $\pi^+$ the distribution should be
modified the opposite way, because it feels a repulsive force. 
To calculate these effects quantitatively 
requires very complicated many-body calculations, including 
Coulomb forces. This is extremely difficult and in order to make a first 
estimate of the effects one might expect the following
simple scaling assumption to be reasonable. 
That is, the change in the pion distributions are assumed to be given by 
\begin{eqnarray}
\delta f_{\pi^+/A}(y) &=& 
 2 \alpha_{\pi^+/A} (1+\beta_{\pi^+/A}) 
f_{\pi^0N}((1+\beta_{\pi^+/A})y) \equiv  2 \alpha_{\pi^+/A} 
f_{\pi^+/A}^*(y) , \label{dpipa} \\
\delta f_{\pi^-/A}(y) &=& 
 2 \alpha_{\pi^-/A} (1-\beta_{\pi^-/A}) 
f_{\pi^0N}((1-\beta_{\pi^-/A})y) \equiv  2 \alpha_{\pi^-/A} 
f_{\pi^-/A}^*(y) , \label{dpima} \\
\delta f_{\pi^+/A'}(y) &=& 
 2 \alpha_{\pi^+/A'} f_{\pi^0N}(y) , \label{dpipap} \\
\delta f_{\pi^-/A'}(y) &=& 
 2 \alpha_{\pi^-/A'} (1-\beta_{\pi^-/A'})
f_{\pi^0N}((1-\beta_{\pi^-/A'})y) \equiv  2 \alpha_{\pi^-/A'} 
f_{\pi^-/A'}^*(y) , \label{dpimap} 
\end{eqnarray}
where $\alpha_{M/A(A')}$ represents a change caused by the strong interaction 
in a nucleus (for example, Pauli blocking, correlations of random phase 
approximation (RPA), etc~\cite{diep}) and $\beta_{M/A(A')} (> 0)$ 
describes a shift of the distribution because of the Coulomb force. 
Note that the $\pi^+$ in $^3$H does not feel any Coulomb force.  
As pointed out above, we expect that $\beta_{\pi^+/A} :
\beta_{\pi^-/A} : \beta_{\pi^-/A'} = 1 : 3 : 2$ from the point of view of
the strength of the Coulomb force acting on the pion.  
We therefore choose  
$3 \beta_{\pi^+/A} = \beta_{\pi^-/A} = \frac{3}{2} \beta_{\pi^-/A'} =
\beta > 0$ ($\beta$ is assumed to be small).
The new function, $f_{M/A(A')}^*$, is normalized as 
\begin{equation}
\int _0^1 dy f_{M/A(A')}^*(y) = \langle f_{\pi^0N}\rangle ,  
\label{nfstr}
\end{equation}
where we ignored a tiny quantity stemming from $\int_{1-\beta_{M/A(A')}}^1 dy 
f_{\pi^0N}(y)$.  

Next, we suppose that the average number of pions per nucleon 
in a nucleus is equal to that in the free nucleon -- experimental 
indications are that  
the pion field is not much enhanced in a nucleus~\cite{exp1,dy}.  
The requirement of pion number conservation reduces the number of parameters 
``$\alpha$'' in Eqs.(\ref{dpipa})-(\ref{dpimap}).  We find that in $^3$He,  
$2\alpha_{\pi^+/A} + \alpha_{\pi^-/A} = 0$, while in $^3$H,  
$\alpha_{\pi^+/A'} + 2\alpha_{\pi^-/A'} = 0$.  
Thus, we set  
$2\alpha_{\pi^+/A} = - \alpha_{\pi^-/A} = \alpha_A > 0$ (the $\pi^-$ field 
is suppressed in $^3$He) and 
$2\alpha_{\pi^-/A'} = - \alpha_{\pi^+/A'} = \alpha_{A'} > 0$ 
(the $\pi^+$ field is suppressed in $^3$H).  Furthermore, 
since $\alpha_{A(A')}$ describes the change of the pion 
field because of the strong interaction, we can set $\alpha_A = \alpha_{A'} 
= \alpha$ (isospin is a good symmetry in this case).  
This leaves just two parameters, $\alpha$ and $\beta$.  

\begin{figure}
\begin{center}
\epsfig{file=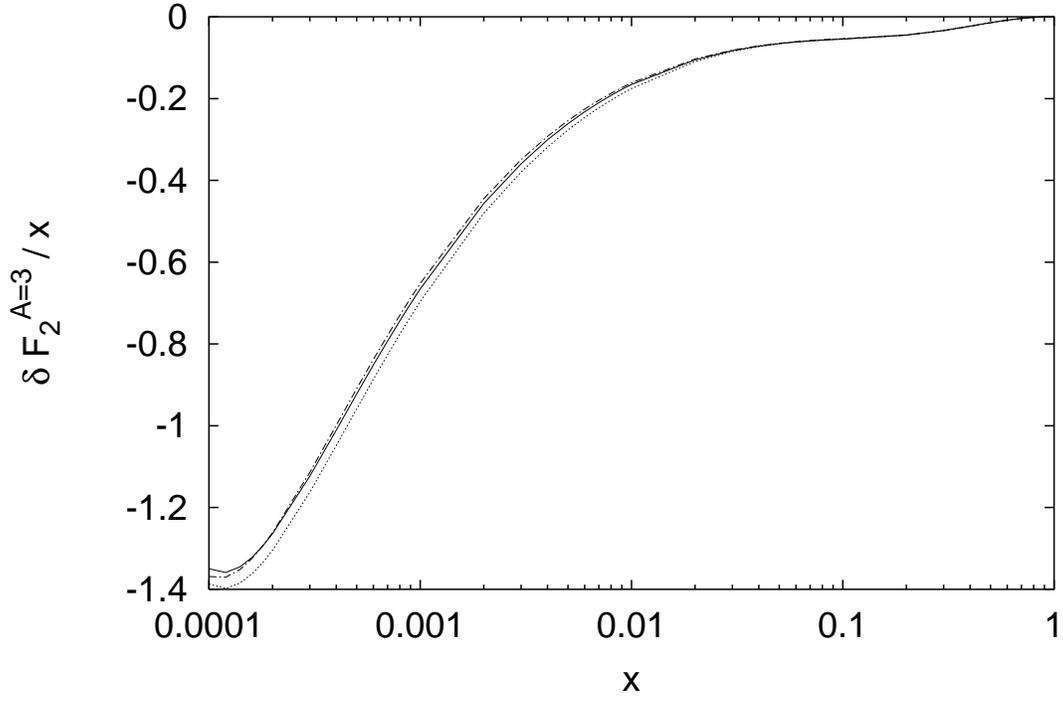,height=10cm}
\caption{
$\delta F_2^{A=3} / x$ vs $x$ ($\alpha = 0.1$).
The dotted, solid and dot-dashed curves show
the results with $\beta=0, 0.1, 0.2$, respectively.
}
\label{fig:diff}
\end{center}
\end{figure}

We should note here that even if $\alpha = 0$ the Coulomb effect would 
modify the proton and neutron structure functions in the nucleus. 
Such a case could be described by replacing the pion distribution in the 
free nucleon structure function, $f_{MB/N}$, 
in Eqs.(\ref{pinA2})-(\ref{ninAp2}) with $f_{M/A(A')}^*$. 
However, we expect that by itself the Coulomb effect on the structure function 
should be quite small (see below Eq.(\ref{dF}) and Fig.~\ref{fig:diff}). 
We therefore neglect the Coulomb effect on $F_2^{p(n)}$ in $F_2^{p(n)/A(A')}$. 
Equations (\ref{pinA2})-(\ref{ninAp2}) then give 
\begin{eqnarray}
\delta F_2^{p/^3He} &\equiv& F_2^{p/^3He} - F_2^p  
= - \alpha \langle f_{\pi^0N} \rangle {\tilde F}_2^p
+ \alpha f_{\pi^+/^3He}^* \otimes [{\tilde F}_2^n +  F_2^{\pi^+}] , 
\label{dpinA} \\
\delta F_2^{n/^3He} &\equiv& F_2^{n/^3He} - F_2^n  
= 2 \alpha \langle f_{\pi^0N} \rangle {\tilde F}_2^n
- 2 \alpha f_{\pi^-/^3He}^* \otimes [{\tilde F}_2^p +  F_2^{\pi^-}] , 
\label{dninA} \\
\delta F_2^{p/^3H} &\equiv& F_2^{p/^3H} - F_2^p  
= 2 \alpha \langle f_{\pi^0N} \rangle {\tilde F}_2^p
- 2 \alpha f_{\pi^0N} \otimes [{\tilde F}_2^n +  F_2^{\pi^+}] , 
\label{dpinAp} \\
\delta F_2^{n/^3H} &\equiv& F_2^{n/^3H} - F_2^n  
= - \alpha \langle f_{\pi^0N} \rangle {\tilde F}_2^n
+ \alpha f_{\pi^-/^3H}^* \otimes [{\tilde F}_2^p +  F_2^{\pi^-}] , 
\label{dninAp}
\end{eqnarray}
where,     
\begin{eqnarray}
f_{\pi^+/^3He}^*(y) &=& \left( 1 +\frac{1}{3}\beta \right)
f_{\pi^0N}\left(\left( 1 +\frac{1}{3}\beta \right) y \right) ,
\label{ff1} \\
f_{\pi^-/^3He}^*(y) &=& ( 1 -\beta )
f_{\pi^0N}((1-\beta)y) ,
\label{ff2} \\
f_{\pi^-/^3H}^*(y) &=& \left( 1 -\frac{2}{3}\beta \right)
f_{\pi^0N}\left(\left( 1 -\frac{2}{3}\beta \right) y \right) .
\label{ff3}
\end{eqnarray}

\begin{figure}
\begin{center}
\epsfig{file=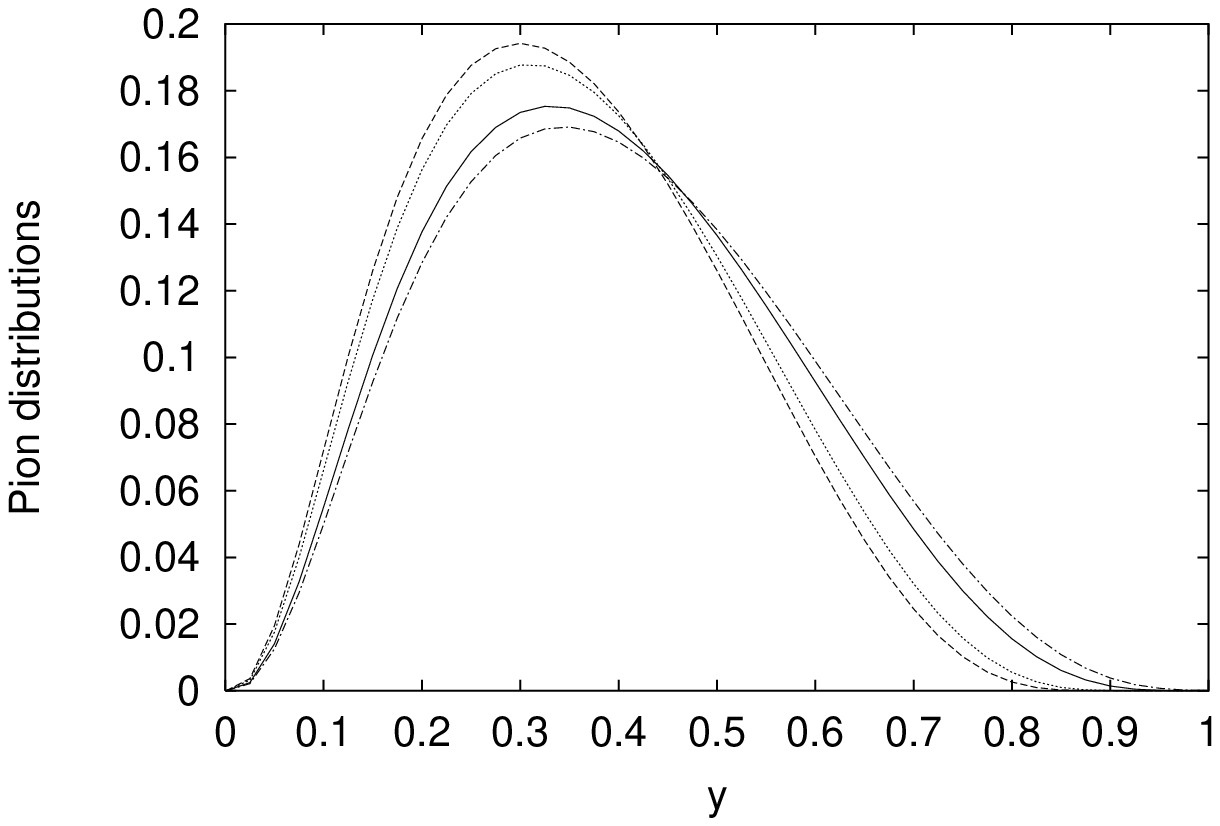,height=10cm}
\caption{
Pion distributions ($\beta = 0.1$ and $\Lambda = 1$ GeV).
The dotted, solid, dot-dashed and dashed curves are for
$f_{\pi^0N}, f_{\pi^-/^3H}^*, f_{\pi^-/^3He}^*$ and
$f_{\pi^+/^3He}^*$, respectively.
}
\label{fig:dist}
\end{center}
\end{figure}

In Fig.~\ref{fig:dist} the pion distributions provided by the
scaling assumption are presented taking $\Lambda = 1$ GeV (see also below 
Eq.(\ref{dsg3})).  As an example, we choose $\beta = 0.1$, which means that, 
for instance, the wave function of the $\pi^-$ in $^3$He shrinks by about 
$10\%$ in coordinate space because of the Coulomb force. 
The negative pion distribution carries somewhat higher momentum, while the 
positive one shifts toward lower $y$, compared with $f_{\pi^0N}$.
Taking the non-singlet combination of the structure functions of $^3$He 
($F_2^{^3He}$) and $^3$H ($F_2^{^3H}$) we find
\begin{eqnarray}
F_2^{^3He} - F_2^{^3H} 
&=& (F_2^p - F_2^n) 
- 4 \alpha \langle f_{\pi^0N} \rangle \delta {\tilde F}_2^N  \nonumber \\ 
&-& 2 \alpha [ f_{\pi^-/^3H}^* + f_{\pi^-/^3He}^* ] \otimes {\tilde F}_2^p 
+ 2 \alpha [ f_{\pi^+/^3He}^* + f_{\pi^0N} ] \otimes {\tilde F}_2^n 
\nonumber \\
&-& 2 \alpha [ f_{\pi^-/^3H}^* \otimes F_2^{\pi^-} + f_{\pi^-/^3He}^* 
\otimes F_2^{\pi^-} -  f_{\pi^+/^3He}^* \otimes F_2^{\pi^+} 
- f_{\pi^0N} \otimes F_2^{\pi^+}] , 
\label{diffin}
\end{eqnarray}
where 
\begin{equation}
\delta{\tilde F}_2^N(x) = {\tilde F}_2^p(x) - {\tilde F}_2^n(x)
= \frac{1}{3} x [ {\tilde u}_v(x) - {\tilde d}_v(x) ] ,
\label{dF}
\end{equation}
with ${\tilde u}_v({\tilde d}_v)$ the valence $u(d)$ distribution in
the bare proton.  

Figure~\ref{fig:diff} illustrates the nuclear 
and Coulomb effects on the non-singlet structure function of the $A=3$ 
system, which is given by $\delta F_2^{A=3} = (F_2^{^3He} - F_2^{^3H}) - 
(F_2^p - F_2^n)$.  For the numerical calculations we have chosen 
(at $Q^2=$ 4 GeV$^2$)~\cite{ericson}
\begin{eqnarray}
x {\tilde u}_v(x) = 0.65452 \times x^{0.38} (1-x)^{2.49} (1+10.5x) ,
\label{uv} \\
x {\tilde d}_v(x) = 0.028660 \times x^{0.07} (1-x)^{4.63} (1+150x),
\label{dv}
\end{eqnarray}
and $F_2^{\pi^+}(x) = F_2^{\pi^-}(x) = 0.98863 \times 
x^{0.61}(1-x)^{1.02}$~\cite{pion2}. 
Clearly the effect of the Coulomb distortion 
is quite small in the region $x > 10^{-4}$, even if 
we choose $\beta = 0.2$. (We have checked that the 
Coulomb effect on $\delta F_2^{N/A(A')}$ individually is also small.) 

The nuclear Gottfried integral, $I_G^{A,A'}(z)$, is defined by~\cite{saito}  
\begin{equation}
I_G^{A,A'}(z) = \frac{1}{Y} \int_z^A \frac{dx}{x}  
[F_2^A(x) - F_2^{A'}(x)] = I_G^N(z) + \delta I_G^{A,A'}(z) , 
\label{gs1}
\end{equation}
with $Y (= Z - N)$ the number of excess protons in $A$ and 
$I_G^N(z)$ the Gottfried integral for the free nucleon. The nuclear 
effect is described by the second term on the r.h.s of 
Eq.(\ref{gs1}) 
\begin{equation}
\delta I_G^{A,A'}(z) = \int_z^A \frac{dx}{x} \delta F_2^{A,A'}(x) , 
\label{gs2}
\end{equation}
where $\delta F_2^{A,A'} = \frac{1}{Y}(F_2^A - F_2^{A'}) - 
(F_2^p - F_2^n)$. 
In the case of the $A=3$ system $\delta F_2^{A,A'}$ is given by 
$\delta F_2^{A=3}$. As we have already discussed in Ref.~\cite{guzey}, 
the Gottfried integral is generally divergent when the effect of 
charge symmetry breaking is included,  
even for the free proton and neutron~\cite{guzey}. 

If we set $\beta = 0$, $\delta F_2^{A=3}$ reads 
\begin{equation}
\delta F_2^{A=3} = 
- 4 \alpha \left[ \langle f_{\pi^0N} \rangle 
+ f_{\pi^0N} \otimes \right] \delta {\tilde F}_2^N . 
\label{diffin2}
\end{equation}
Since the effects of the nuclear binding and shadowing 
are ignored in the calculation for the time being, the change in the 
Gottfried integral is convergent and it is given by 
\begin{equation}
\delta I_G^{^3He,^3H}(0) = - \frac{8}{3} \alpha 
\langle f_{\pi^0N} \rangle .  
\label{dsg3}
\end{equation}
With the cut-off-mass $\Lambda = 1$ GeV in $f_{\pi^0N}$, we find 
$\langle f_{\pi^0N} \rangle = 0.083$. The Gottfried 
integral for the free nucleon is then estimated to be  
$I_G^N(0) = \frac{1}{3}(1 -4 \langle f_{\pi^0N} \rangle ) = 0.223$, which 
is consistent with the measured value ($0.235 \pm 0.026$)~\cite{review1}.  
We can see that the modification of the pion field enhances the flavor 
asymmetry in a pair of mirror nuclei and hence it reduces 
the Gottfried integral. 
If $\alpha = 0.05 (0.1) [0.2]$ (and $\beta = 0$), $\delta I_G^{^3He,^3H}(0) 
= -0.0111 (-0.0221) [-0.0443]$, corresponding to a reduction of the 
Gottfried sum by about $ 5 (9) [20] \%$ from the free value.  

Next we consider the nuclear binding and shadowing effects. 
(For recent reviews see Ref.~\cite{review2}.)  
In Ref.~\cite{guzey} we studied shadowing and anti-shadowing 
corrections to the flavor non-singlet structure function 
using the Gribov-Glauber multiple scattering formalism. 
We found that the non-singlet structure function is enhanced at 
small $x$ by nuclear shadowing, increasing the nuclear Gottfried 
integral ($z$ is chosen to be $10^{-4}$ in Eq.(\ref{gs1})) by between  
$15$ and $41\%$.  
The enhancement of the non-singlet structure function is caused by 
the difference between the density distributions of $^3$He and $^3$H.  

In the shadowing region the structure functions of $^3$He and $^3$H are 
given by~\cite{guzey}
\begin{eqnarray}
F_2^{^3He} &=& 2 F_2^p + F_2^n - (2.5 f_{^3He} - g_{^3He}) F_2^p 
- 0.5 f_{^3He} F_2^n , \label{hesh} \\
F_2^{^3H} &=& F_2^p + 2 F_2^n - (2.5 f_{^3H} - g_{^3H}) F_2^n 
- 0.5 f_{^3H} F_2^p ,
\label{hsh} 
\end{eqnarray}
where $f_{^3He(^3H)}$ and $g_{^3He(^3H)}$, respectively, describe the single 
and double rescattering processes in $^3$He ($^3$H), which depend on the 
nuclear density distribution. It is also necessary to include  
anti-shadowing in order to produce the structure function around 
$x \sim 0.1$ (using the baryon number and momentum sum rules). 
Detailed discussions can be found in Ref.~\cite{guzey}. 

Replacing the structure functions of the free proton and neutron in 
Eqs.(\ref{hesh}) and (\ref{hsh}) by $F_2^{N/A(A')}$ 
(given by Eqs.(\ref{pinA2})-(\ref{ninAp2})), we can calculate the 
non-singlet structure function of $^3$He and $^3$H, including both 
the modification of the pion fields and nuclear shadowing.  
(This means that the effect of the pion cloud modifies the nucleon sea 
quarks only.) Note 
that this simple replacement is an approximation made in order 
to see the effect of the 
change of the pion fields. To treat  
this problem rigorously it would be 
necessary to construct a model where pions are 
handled consistently and contribute to both shadowing and
anti-shadowing~\cite{Tonymec}, which goes beyond the scope of 
the present work.

In the calculation of nuclear shadowing, the ground-state wave functions 
of $^3$He and $^3$H are assumed to be given by gaussian functions in 
coordinate space~\cite{guzey}, $|\Psi|^2 \propto
\exp[-{\vec r}^{\,2}/(2b)]$, where the parameter $b$ determines the 
correct matter
radius of the nucleus. We take $b = 40.59 (30.06)$ GeV$^{-2}$, which 
produces the matter radius of 1.769 (1.524) fm for $^3$He ($^3$H). 
In Ref.~\cite{guzey} the effective cross section, $\sigma_{eff}$, was 
used to describe the 
interaction between the hadronic components in the virtual photon and 
the nuclear target.  Here we take two models for 
$\sigma_{eff}$: the first (case 1) from Frankfurt and 
Strikman~\cite{fs}, and the second (case 2)  
the two-phase model of Ref.~\cite{wt}. 
At large $x (> 0.2)$ 
we used the structure functions of $^3$He and $^3$H, obtained as a solution 
of the Faddeev equations for three body system~\cite{saito,bissey}. 
Since the contribution of the nuclear binding and Fermi motion effects to 
the non-singlet combinations of the structure functions of 
$^3$He and $^3$H is small, 
we make an approximation that the binding and Fermi motion 
effect of Refs.~\cite{saito,bissey} 
and the pion cloud effect, discussed in the 
present work, are not correlated and, hence, contribute additively to 
$\delta F_2^{A=3}$. 

\begin{figure}
\begin{center}
\epsfig{file=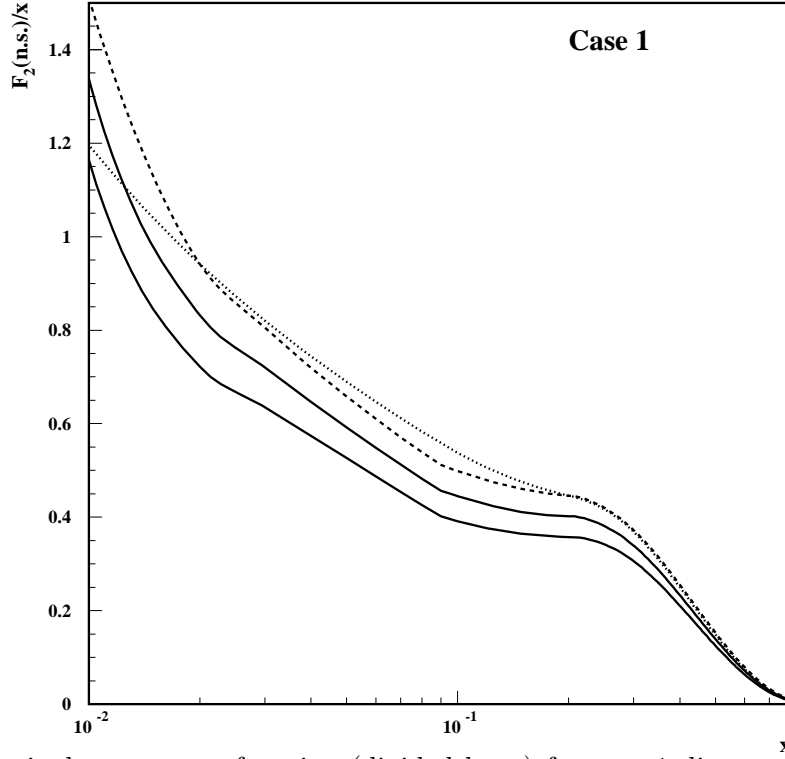,height=10cm}
\caption{
Non-singlet structure function (divided by $x$) for case 1 discussed 
in the text. 
The dotted curve shows the non-singlet structure function 
for the free nucleon, while the dashed curve presents the result for 
the $A=3$ system without any change of the pion fields. 
The upper (lower) solid curve is for the full calculation with 
$\alpha = 0.1 (0.2)$. 
}
\label{fig:fin1}
\end{center}
\end{figure}

\begin{figure}
\begin{center}
\epsfig{file=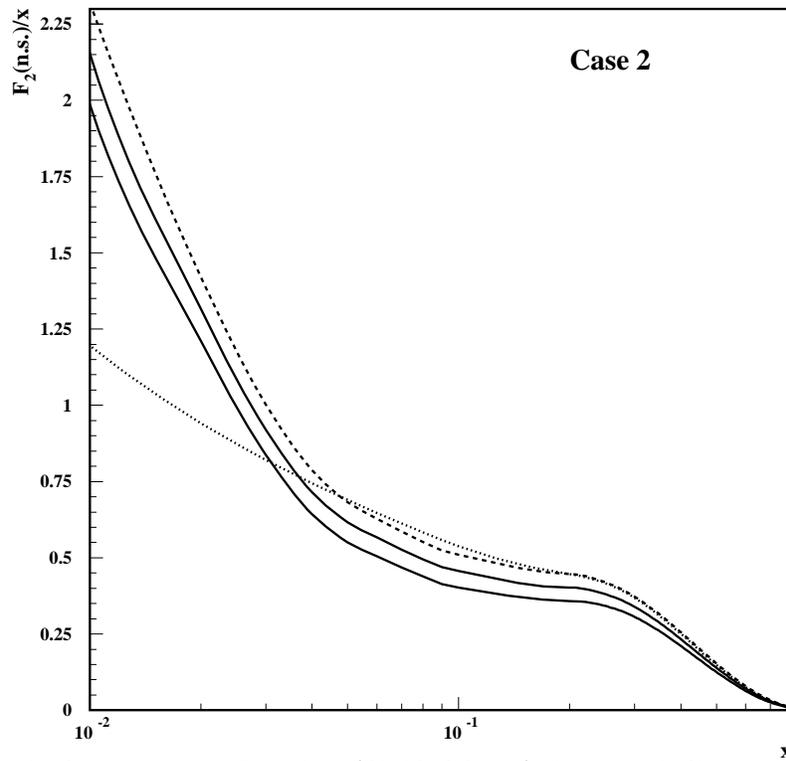,height=10cm}
\caption{
Non-singlet structure function (divided by $x$) for case 2 discussed in the 
text. The curves are labelled as in Fig.~\protect\ref{fig:fin1}. 
}
\label{fig:fin2}
\end{center}
\end{figure}

We present our main results in Figs.~\ref{fig:fin1} and \ref{fig:fin2} 
(the calculations were performed at $Q^2 = 4$ GeV$^2$).  To treat parton 
densities in the free proton and neutron realistically we used the CTEQ5L 
parametrization~\cite{cteq}. Since we know that the Coulomb effect is 
small in the region $x > 10^{-4}$, we set $\beta=0$ and show only 
the dependence of the flavor non-singlet structure function on $\alpha$. 

As expected, the change of the pion fields leads to a considerable 
suppression of the non-singlet structure function in a nucleus. 
However, at very small $x$ 
(typically $\sim 10^{-3}-10^{-4}$) the reduction is 
not large compared with the enhancement caused by shadowing.  
(In the figures we present our results only for the range 
$x \in [10^{-2},0.8]$ in order 
to see the difference among the curves clearly. Note that the 
non-singlet structure function (divided by $x$) is of order $200 - 300$ at 
$x = 10^{-4}$.) 
When the shadowing is turned off the difference of the proton and neutron 
structure functions gives the main contribution to the non-singlet structure 
function of the nucleus. In this case the change of the pion fields gives 
a sizable contribution at very small $x$ (e.g., about $20\%$ at 
$x = 10^{-3}$).  On the contrary, when the shadowing is switched on the 
main contribution to the non-singlet structure function at small $x$ is 
given by a term 
proportional to $(F_2^p + F_2^n)\times(f_{^3He} - f_{^3H})$, which 
is much larger than the effect of the change of the pion fields -- 
the pion effect is at most $5\%$ at $x = 10^{-3}$. 

We can estimate the nuclear Gottfried integral, $I_G^{^3He,^3H}(10^{-4})$, 
defined by Eq.(\ref{gs1}).  If $\alpha$ is set to be $0$ (no change in the 
pion fields) we find $I_G^{^3He,^3H}(10^{-4}) = 0.2953 (0.3395)$ for case 1 
(2) -- note that the CTEQ5L fit gives $I_G^N(10^{-4}) = 0.2403$. 
When $\alpha = 0.1 (0.2)$ we obtain $I_G^{^3He,^3H}(10^{-4}) = 0.2699 
(0.2444)$ for case 1, while it is $0.3142 (0.2829)$ for 
case 2. Therefore, we expect that the change of the pion fields in the 
$A=3$ system might lead to a reduction of the Gottfried integral by about 
$10\%$, compared with the value for the case where $\alpha = 0$. 

We here give some comments: \\
\noindent (1) Using the scaling assumption the pion distributions 
are shifted in the present calculation.  
The momentum fraction carried by pions is thus different from 
that in the free nucleon. For example, if 
$0 < \beta \ll 1$, we find $\langle y f_{\pi^-/^3He}^*(y) \rangle \simeq 
(1 + \beta) \langle y f_{\pi^0N}(y) \rangle$.  Thus,  
in $^3$He ($^3$H) the ratio of the momentum fraction carried by pions to 
that in the case where $\beta = 0$ increases by about $\beta/9$ ($4\beta/9$). 
The change of the pion momentum fraction would lead to changes of the 
momentum fractions carried by the nucleons and other mesons. In such a 
case one would have to construct a model 
where the momentum fractions are balanced. Note however, that 
when we set $\beta = 0$ there is no change of the pion momentum 
fraction in a nucleus by virtue of the scaling 
assumption and pion number conservation.  

\noindent (2) We have treated the change of the pion 
fields in the nuclear medium semi-quantitatively. In principle, 
such a modification should be attributed to various correlation phenomena 
in a nucleus like Pauli blocking, exchange currents, short-range 
correlations etc. 
Melnitchouk and Thomas~\cite{mt2} have reanalysed the nuclear 
shadowing effect on 
the deuteron structure function, including meson exchange currents, to 
extract the neutron structure function, and studied the Gottfried 
sum rule for the free nucleon. 
It should be possible to do such calculations for the three body 
system in the future. For a pair of mirror nuclei larger than the three body 
system, it would become important to consider Fock states including 
multi-pions, $\Delta$ isobars, other mesons and so on.  That is also a very 
intriguing problem. 

In summary, we have estimated the effect of
the medium modification of
the pion fields on the flavor non-singlet structure function of the
lightest mirror nuclei.  We have found that the change of the pion fields
produces a considerable suppression of the non-singlet structure function,
and that the Gottfried integral is correspondingly reduced. 
In general, charge symmetry is broken and 
the Gottfried integral is divergent~\cite{guzey}.  However, 
the $x$-dependence of the flavor non-singlet structure function of a pair 
of mirror nuclei would provide significant information on 
phenomena involving non-pQCD dynamics (such as the pion cloud) 
in the nuclear medium. 
Experiments on deep-inelastic scattering off various mirror nuclei should be 
possible in the future~\cite{future}.  If we could vary 
the atomic number $A$ and the difference between the proton and neutron 
numbers $Y$ independently in measuring the nuclear structure functions, 
it would stimulate a great deal of work which should lead to 
new information on the dynamics of nuclear systems. 

\vspace{0.5cm}
We would like to thank F. Bissey for providing us with the results 
for the structure functions of $^3$He and $^3$H in the large Bjorken 
$x$ region. This work was supported by the 
Australian Research Council and Adelaide University.


\newpage
\newpage
\newpage
\newpage

\end{document}